\documentclass[intlimits,twoside,a4paper]{article}
\usepackage{amsmath,amssymb}
\usepackage{graphicx}

\usepackage{algorithm}
\usepackage{algorithmic}

\usepackage[T2A]{fontenc}
\usepackage[cp1251]{inputenc}
\usepackage{cmpj2}

\issue{2014}{17}{3}{33401}
\doinumber{10.5488/CMP.17.33401}

\title[Self-organization of adatom adsorption structure at interaction with the tip of dynamic force microscope]%
{Self-organization of adatom adsorption structure at interaction with tip of dynamic force microscope}
\authorcopyright{A.V. Khomenko, 2014}
\author{A.V. Khomenko\refaddr{label1,label2}}
\addresses{
\addr{label1} Department of Complex Systems Modelling, Sumy State University, \\2 Rimskii-Korsakov St., 40007 Sumy, Ukraine
\addr{label2} Peter Gr\"unberg Institut-1, Forschungszentrum-J\"ulich, 52425 J\"ulich, Germany
}

\date{Received April 5, 2014, in final form June 04, 2014}
\begin{document}

\maketitle

\begin{abstract}
The formation of an adatom adsorption structure in dynamic force microscopy experiment is shown
as a result of the spontaneous appearance of shear strain caused by external supercritical heating.
This transition is described by the Kelvin-Voigt equation for a viscoelastic medium, the relaxation
Landau-Khalatnikov equation for shear stress,  and the relaxation equation for temperature.
It is shown that these equations formally coincide with the synergetic Lorenz system, where
the shear strain acts as the order parameter, the conjugate field is reduced to the stress,
and the temperature is the control parameter. Within the adiabatic approximation, the steady-state
values of these quantities are found. Taking into account the sample shear modulus vs strain dependence,
the formation of the adatom adsorption configuration is described as the first-order transition.
The critical temperature of the tip linearly increases with the growth of the effective value of the sample
shear modulus and  decreases with the growth of its typical value.
\keywords phase transition, rheology, plasticity, strain, stress, atomic force microscopy
\pacs 46.55.+d, 
    64.60.-i, 
    61.72.Hh, 
    62.20.F-, 
    62.20.Qp, 
    68.37.Ps 

\end{abstract}

\section{Introduction}\label{sec:level1}

Nowadays, due to large scientific and practical importance, the phenomena taking place on the sample surface at interaction with the tip of a  dynamic force microscope, e.g., atomic force microscope (AFM) and friction force microscope, attract more and more attention (see the reviews in \cite{book_Gnecco, Garcia_book, SSR_Perez, JPD_rev, RMP_Hofer, RMP_Giessibl} and the literature cited therein). Particularly, the experimental and theoretical data are obtained on structural instabilities, phase transformations, plastic dislocation, neck and adatom structures formation \cite{Pogrebnjak2009,PRB_2006,PRL_2004,Roy_plast,PRL_2006,Khome2010,SS_2010,Khomenko2010jpc, Pogrebnyak:2012,SMALL_2012,Nanotechnology_2011,Adv_Mat_2011,Nanotechnology_2010,PhysRevB.79.125431}. These processes are characterized by hysteresis of dependencies of adhesion force and potential energy surface on the tip-surface distance \cite{PRL_2006, TPL, PRL_2005, SMALL_2012, PhysRevB.85.235412, Lange2012} and by hysteresis of the  sample stress vs strain curve \cite{PRL_2004_sig}.

Since the nature of such phenomena remains poorly understood, the basic goal of the present study is the construction of a qualitative nonlinear model \cite{Haken,Landau,kin,zhetph,KhYu,pla} describing the hysteresis processes which occur on the germanium surface during interaction with the AFM tip \cite{PRL_2006}. Here, the macroscopic continuum mechanics models \cite{voigt,upr} are supposed to be still applicable to the atomic length-scales, where discrete atomistic interactions become significant \cite{SSR_Perez,SMALL_2012,Adv_Mat_2011,Mate_1995,prd}. However, a total explanation of the studied macroscopic phenomena requires a consideration of microscopic processes.  Phenomenological description used here makes it possible to connect the parameters of microscopic theories with macroscopic measurements. However, this is a separate independent problem that is hard to solve just now. In the presented approach, the formation conditions of the adatom adsorption structure are defined on the semiconductor surface due to both thermal and deformation effects. The total set of freedom degrees is considered as equivalent variables. The adatom configuration formation is described analytically as a result of self-organization caused by the positive feedback of shear strain and temperature on shear stress on the one hand, as well as the negative feedback of shear strain and stress on temperature on the other hand. This study is based on the assumption that stress relaxation time diverges because the shear modulus vanishes at the point of transition.

The paper is organized as follows. In section~\ref{sec:level2}
the self-consistent Lorenz system of the governed equations is written for approximation of semiconductor characterized by heat conductivity.
The adatom structure formation is shown in section~\ref{sec:level3} to be supercritical in character (is of the second order) when the effective shear modulus of the germanium does not depend on the strain value; it then transforms to a subcritical mode with this dependence appearance (section~\ref{sec:level4}).
In these sections, the steady-state values of shear strain and stress, as well as temperature are also determined within adiabatic approximation. Using such a limit, a synergetic potential is obtained, that is the analog of a thermodynamic potential, from basic evolution equations. Section~\ref{sec:level5} contains short conclusions.

\section{Basic equations}\label{sec:level2}

Let us start with the supposition that the relaxation behavior of the shear component $\varepsilon$ of the strain tensor in a semiconductor is governed by the Kelvin-Voigt equation \cite{voigt,Ultramicroscopy_2010}
\begin{equation}
\dot {\varepsilon}=-\varepsilon/\tau_{\varepsilon}+\sigma/\eta_{\varepsilon}\,,
\label{7} \end{equation}
where $\tau_{\varepsilon}$ is the Debye relaxation time and
$\eta_{\varepsilon}$ is the effective shear viscosity coefficient.
The second term on the right-hand side describes the flow of a
viscous liquid caused by the corresponding shear component of the stress $\sigma$.
In the steady state, $\dot\varepsilon = 0$, we obtain the Hooke-type expression
$\sigma = G_{\varepsilon}\varepsilon,~ G_{\varepsilon} \equiv \eta_{\varepsilon}/\tau_{\varepsilon}$.

The next assumption of our approach is that the relaxation equation of the sample shear stress $\sigma$ has a form similar to the Landau-Khalatnikov equation \cite{kin, coll, KhYu}:
\begin{equation}
\tau_\sigma\dot\sigma = -\sigma + G(T)\varepsilon .
\label{6} \end{equation}
Here, the first term on the right-hand side describes the relaxation
during time $\tau_{\sigma} \equiv  \eta/G(T)$ determined by the values of the
shear viscosity $\eta$ and modulus $G(T)$ depending on the sample temperature.
In the stationary case $\dot \sigma = 0$, the kinetic
equation (\ref{6}) is transformed into the Hooke's law
\begin{equation} \sigma = G(T)\varepsilon.  \label{6a}
\end{equation}

Note that effective values of viscosity $\eta_\varepsilon \equiv \tau_\varepsilon G_\varepsilon$
and modulus $G_{\varepsilon} \equiv \eta_\varepsilon/\tau_{\varepsilon}$ do not
coincide with the real values $\eta$ and $G(T)$. Physically, such difference is conditioned by
 the Landau-Khalatnikov-type equation~(\ref{6}) being not equivalent to
the Kelvin-Voigt equation (\ref{7}) \cite{voigt, upr, KhYu}. As is known, the values $G_\varepsilon,~\eta,~\eta_\varepsilon$  very weakly depend on the sample temperature $T$, while the real shear modulus $G(T)$ vanishes, when the temperature decreases to $T_{\mathrm{c}}$ \cite{RMP_11,marvan,glass1,glass2,cmp_2006}. Further, the simplest approximate temperature dependencies are used:
$G_{\varepsilon}(T),~\eta(T), \eta_\varepsilon(T) = {\rm const}$,
\begin{equation}
G(T) = G_0\left(T/T_{\mathrm{c}}-1\right), \label{8}
\end{equation}
where $G_{0} \equiv G(T=2T_{\mathrm{c}})$ is the typical value of modulus.

According to the synergetic concept \cite{Haken,zhetph,KhYu,glass1,glass2,physa_soc,jam_pre} to complete the equation system (\ref{7}) and (\ref{6}), which contains the order parameter $\varepsilon$, the conjugate field $\sigma$, and the control parameter $T$, we should deduce a kinetic equation for the temperature.
This equation can be obtained using the basic relationships of elasticity theory
stated in $\S~32$ in \cite{upr}. Thus, it is necessary to start with the continuity
equation for the heat $\delta Q = T\delta S$:
\begin{equation}
T\dot S = - \nabla { \bf q}.  \label{13gl}
\end{equation} Here, the heat
current is given by the Onsager equation
\begin{equation} {\bf q} =
-\kappa \nabla T, \label{14gl}
\end{equation}
where $\kappa$ is the heat conductivity.  In the elementary case of the thermoelastic
stress, the entropy
\begin {equation} S = S_0 (T) + K\alpha\varepsilon^0
\label{15gl}\end{equation}
consists of the purely thermodynamic component $S_0$ and the dilatation:
\begin{equation}
\widehat \varepsilon^0 = \varepsilon^0\widehat I,
\qquad \varepsilon^0\equiv\alpha\left (T - T_0 \right), \label{5gl}
\end{equation}
where $\alpha$ is the thermal expansion coefficient,
$T_0$ is the equilibrium temperature, $\widehat I$ is the unit
tensor and $K$ is the compression modulus (see $\S~6$ in \cite{upr}). In the considered situation, we should transfer from the dilatational component $K\alpha \varepsilon^0$ to the elastic energy $-\sigma\varepsilon /T$ of the shear component divided by temperature (here, the minus sign takes into account the connection
$T\delta S = p\delta V \Rightarrow -
\sigma\delta\varepsilon$ at $S_0$ = const, which is caused by the opposite choice of the
pressure $p$ and the stress $\sigma$ signs). As a result, equation~(\ref{13gl}) has the form
\begin{equation} T\dot S_0 (T)
-\sigma\dot\varepsilon = \kappa\nabla^2T.
\label{16gl} \end{equation}
Taking into account the approximation $(\kappa /l^2) (T_{\mathrm{e}}-T)\approx \kappa\nabla^2T$
($l$ is the scale of heat conductivity, $T_{\mathrm{e}}$ is the AFM tip temperature) and the definition of heat capacity
$c_{\mathrm{p}}{=}T{\rm d}S_0/{\rm d}T$, equation~(\ref{16gl}) assumes the form:
\begin{equation}
c_{\mathrm{p}}\dot T = \frac{\kappa}{l^2}(T_{\mathrm{e}}-T) + \sigma\dot\varepsilon. \label{17gl}
\end{equation}
Substituting the expression for the $\dot\varepsilon$ from equation~(\ref{7}) we obtain the term
$\sigma^2/\eta_\varepsilon$. It describes the dissipative heating of a viscous liquid flowing under the effect of the stress $\sigma$ that can be neglected in the case under consideration. On the other hand, the process of an AFM tip moving into contact with the surface has the following
peculiarity. It is necessary to consider the thermal effect of the tip
whose value $T_{\mathrm{e}}$ is not reduced to the Onsager component and is fixed by external conditions.
In view of these circumstances, the square contribution of the
stress is supposed to be included in $T_{\mathrm{e}}$. The obvious account of this term leads to a
significant complication of the subsequent analysis, though it results
in a renormalization of the quantities. Therefore, component $T_{\mathrm{e}}$ in equation~(\ref{17gl}) is assumed to be constant for our further consideration.

It is convenient to introduce the following measure units:
\begin{equation}
\sigma_{s}=\left(c_{\mathrm{p}}\eta_\varepsilon T_{\mathrm{c}}/\tau _{\mathrm{T}}\right)^{1/2},
\qquad \varepsilon_{s}= \sigma_{s}/G_\varepsilon\,,\qquad T_{\mathrm{c}}\,,
\label{10} \end{equation}
for the variables $\sigma$, $\varepsilon $, $T$, respectively
($\tau_{\mathrm{T}}\equiv l^2 c_{\mathrm{p}} / \kappa$ is the time of heat
conductivity). Then, the basic equations (\ref{7}), (\ref{6}), and (\ref{17gl}) take the form:
\begin{eqnarray}
&&\tau_{\varepsilon }\dot{\varepsilon}=-\varepsilon + \sigma ,
\label{11} \\
&&\tau_{\sigma}\dot{\sigma}=-\sigma +g(T-1)\varepsilon , \label{12} \\
&&\tau _{\mathrm{T}}\dot{T}=(T_{\mathrm{e}}-T) - \sigma \varepsilon,
\label{13} \end{eqnarray} where the constant
\begin{eqnarray} g=\frac{G_0}{G_\varepsilon} \label{14} \end{eqnarray} is introduced. Equations (\ref{11})--(\ref{13}) have a form similar to the Lorenz scheme \cite{Haken} which allows us to describe the thermodynamic phase and the kinetic transitions \cite{zhetph,KhYu,glass1,glass2,physa_soc,jam_pre,TEPH-2010_1}.

\section{Continuous transition}\label{sec:level3}

In general the system (\ref{11})--(\ref{13}) cannot be solved analytically. Therefore, we use the following adiabatic approximation:
\begin{equation}
\tau_{\sigma} \ll \tau_{\varepsilon}\,,\qquad \tau_{\mathrm{T}} \ll \tau _{\varepsilon}\,.
\label{15} \end{equation}
This implies that in the course of the matter evolution, the stress $\sigma(t)$
and the temperature $T(t)$ follow the variation of the strain $\varepsilon(t)$.
The first of these inequalities is fulfilled because it contains the macroscopic time $\tau_\varepsilon$
and the microscopic Debye time $\tau_{\sigma}\approx a/c {\sim} 10^{-12}$ s,
where $a\sim 1$~nm is the lattice constant or the intermolecular distance
and $c\sim 10^3$~m/s is the sound velocity. The second condition (\ref{15}) can be reduced to the form
\begin{equation} l \ll L,
\label{e} \end{equation}
where the maximal value of the characteristic length of the heat conductivity
\begin {equation}
L = \sqrt{\chi\nu_\varepsilon \over c_\varepsilon^2}\,, \label{f} \end{equation}
the thermometric conductivity $\chi \equiv \kappa /c_{\mathrm{p}}$, the effective kinematic viscosity $\nu_\varepsilon \equiv \eta_\varepsilon / \rho$ and the sound velocity $c_\varepsilon \equiv(G_\varepsilon/\rho)^{1/2}$
are introduced ($\rho$ is the medium density). Then, we can put the left-hand sides of equations~(\ref{12}) and (\ref{13}) to be
equal to zero. As a result, the stress $\sigma$ and the temperature $T$ are expressed in terms of the strain $\varepsilon$:
\begin{eqnarray}
&&\sigma = \frac{g\varepsilon \left(T_{\mathrm{e}} -1 \right)}{1+g\varepsilon^{2} }\,, \label{16} \\
&&T=1+\frac{T_{\mathrm{e}} -1}{1+g\varepsilon^{2}}\,.  \label{17}
\end{eqnarray}
In accordance with equation~(\ref{17}), in the important range of values of the parameter $T_{\mathrm{e}}>1$,
the temperature $T$ decreases monotonously with an increasing strain $\varepsilon$ from the value
$T_{\mathrm{e}}$ at $\varepsilon = 0$ to $(T_{\mathrm{e}}+1)/2$ at $\varepsilon =\varepsilon_{\mathrm{m}}\equiv\sqrt{1/g}$. Obviously, this decrease is caused by the negative feedback of the stress and the strain on the temperature in equation~(\ref{13}), which is explained by the Le Chatelier principle for this problem. Really, the reason
for the formation of adatom adsorption structure is the
positive feedback of the strain and the temperature on the stress in equation~(\ref{12}).
Hence, the increase in the temperature should intensify the
self-organization effect. However, according to equation~(\ref{13}), the system
behaves in such a way that the consequence of transition, i.e.,
 the growth of the strain, leads to a decrease in its cause
(i.e., temperature). Equation~(\ref{16}), expressing the stress in terms
of the strain, has a linear form of the Hooke's law at
$\varepsilon \ll \varepsilon_{\mathrm{m}}$ with the effective shear modulus $ G_{\mathrm{ef}} \equiv g\left(T_{\mathrm{e}} -1 \right)$.
At $\varepsilon = \varepsilon_{\mathrm{m}}$, the function $\sigma (\varepsilon)$
has a maximum and at $\varepsilon > \varepsilon_{\mathrm{m}}$ it decreases,
which has no physical meaning. Thus, the constant $\varepsilon_{\mathrm{m}}\equiv\sqrt{1/g}$ gives the maximal strain. An increase in the typical value of the modulus $G_0$ leads to a decrease in the maximal strain $\varepsilon_{\mathrm{m}}$ and to an increase in the effective modulus $G_{\mathrm{ef}} $ whose
value is proportional to the characteristic temperature $T_{\mathrm{e}}$.

Substituting equation~(\ref{16}) into equation~(\ref{11}), we obtain the
Landau-Khalatnikov-type equation \cite{kin,coll,PhysRevE_Metlov,PhysRevLett_Metlov}
\begin{equation}
\tau_{\varepsilon}\dot{\varepsilon}=-\partial V/\partial \varepsilon, \label{18}
\end{equation}
where the synergetic potential has the form
\begin{equation}
V=\frac{1}{2}\left[\varepsilon^{2}+\left(1-T_{\mathrm{e}}\right)\ln \left(1+g\varepsilon^{2}\right)\right]. \label{19}
\end{equation}
At a steady state, the condition $\dot{\varepsilon}=0$ is fulfilled
and the potential (\ref{19}) acquires a minimum. When the
temperature $T_{\mathrm{e}}$ becomes smaller than the critical value
\begin{equation}
T_{\mathrm{c}0}=1+g^{-1};\qquad g\equiv G_{0}/G_\varepsilon < 1,\qquad G_\varepsilon \equiv \eta_\varepsilon/
\tau _{\varepsilon}\,, \label{20}
\end{equation}
this minimum corresponds to $\varepsilon=0$, i.e., the adatom adsorption structure is not realized. In the reverse case $T_{\mathrm{e}}>T_{\mathrm{c}0}$, the stationary shear strain has the nonzero value
\begin{equation}
\varepsilon_{0}=\left[T_{\mathrm{e}}-(1+g^{-1})\right]^{1/2} \label{21}
\end{equation}
which increases with $T_{\mathrm{e}}$ growth in accordance with the root law. This causes the formation of the adatom configuration. Equations~(\ref{16}) and (\ref{17}) give the stationary values of stress and temperature:
\begin{equation}
\sigma_{0}=\varepsilon_{0}\,, \qquad T_{0}=1+g^{-1}. \label{22}
\end{equation}
Note that, on the one hand, the steady temperature $T_{0}$
coincides with the critical value (\ref{20}) and, on the other hand,
its value differs from the temperature $T_{\mathrm{e}}$.
Since $T_{\mathrm{c}0}$ is the minimal temperature at which the formation of the adatom adsorption structures can be observed, the above implies that the negative feedback
of the stress $\sigma$ and the strain $\varepsilon$ on the
temperature $T$ [see the last term on the right-hand side of equation~(\ref{13})]
decreases the sample temperature so much that only in the limit does it ensure the
self-organization process. At a steady state, the value of the shear modulus is
\begin{equation}
G_{s} = G_{\varepsilon}\,. \label{23}
\end{equation}

The two cases can be marked out by the parameter $g=G_{0}/G_\varepsilon$.
In the situation $g\gg 1$, meeting the large value of the modulus $G_{0}$,
equations~(\ref{20})--(\ref{22}) take the form
\begin{equation} \varepsilon_{0}=(T_{\mathrm{e}}-1)^{1/2}, \qquad T_{0}=T_{\mathrm{c}0}=1. \label{24}
\end{equation} This corresponds to the ``solid (fragile)'' limit. The opposite case $g\ll 1$ (small modulus $G_{0}$) meets the ``strongly viscous liquid''
\begin{equation} \varepsilon_{0}=\left(T_{\mathrm{e}}-g^{-1}\right)^{1/2},\qquad T_{0}=T_{\mathrm{c}0}=g^{-1}=G_\varepsilon/G_{0}\,.
\label{25} \end{equation}

\section{Effect of deformational defect of modulus}\label{sec:level4}

The Kelvin-Voigt equation (\ref{7}) assumes the use of the idealized
Genki model. For the dependence $\sigma (\varepsilon)$ of the stress
on the strain, this model is described by the Hooke's expression
$\sigma = G_\varepsilon\varepsilon$ at $\varepsilon < \varepsilon_{\mathrm{m}}$ and by the
constant $\sigma_{\mathrm{m}}=G_\varepsilon\varepsilon_{\mathrm{m}}$ at $\varepsilon\geqslant \varepsilon_{\mathrm{m}}$
($\sigma_{\mathrm{m}}$, $\varepsilon_{\mathrm{m}}$ are the maximal stress and strain,
$\sigma > \sigma_{\mathrm{m}}$ results in viscous flow with the
deformation rate $\dot\varepsilon = \left (\sigma - \sigma_{\mathrm{m}}\right)
/\eta_\varepsilon$). Actually, the $\sigma$ vs  $\varepsilon$ dependence curve
has two regions: the first one, Hookean, has a large slope corresponding to the shear modulus $G_\varepsilon$, followed by a more gently sloping
section of the plastic deformation whose tilt is defined by
the hardening factor $\Theta < G_\varepsilon$. Obviously, such a picture means
that the shear modulus, introduced in equation~(\ref{7}), depends on the strain value. Let us use the
simplest approximation \cite{jfd_2007,Khomenko2007165}
\begin{equation} G_\varepsilon (\varepsilon) = \Theta + {G_\varepsilon - \Theta \over 1 +
\left( \varepsilon /\varepsilon_{\mathrm{p}} \right)^2 }\,, \label{47}
\end{equation}
which describes the above mentioned transition of the
elastic deformation mode to the plastic one. It takes place at a
characteristic value of the strain $\varepsilon_{\mathrm{p}}$, which is smaller than $\varepsilon_s$ (otherwise plastic mode is not realized). Note that an expression of the type
equation~(\ref{47}) was originally proposed by Haken \cite{Haken} describing the rigid mode of laser radiation. It is used \cite{zhetph,physa_soc,jam_pre} to describe the first-order phase transition, and equation~(\ref{47}) contained the square of the ratio $\varepsilon /\varepsilon_{\mathrm{p}}$ (so, the
$V$ vs $\varepsilon$ dependencies in \cite{zhetph,physa_soc,jam_pre} and equation~(\ref{51}) have an even form). In
the description of structural phase transitions of a liquid, the third-order
invariants, breaking the specified parity, are present \cite{Landau}. Therefore, in the study~\cite{KhYu,glass1,glass2},
in the approximation (\ref{47}), we used the linear term $\varepsilon /\varepsilon_{\mathrm{p}}$, instead
of the square term $\left (\varepsilon /\varepsilon_{\mathrm{p}}\right)^2$. Obviously, in this
case, the $V$ vs $\varepsilon$ dependence is already uneven.

\begin{figure}[ht]
\centerline{\includegraphics[width=0.3\textwidth]{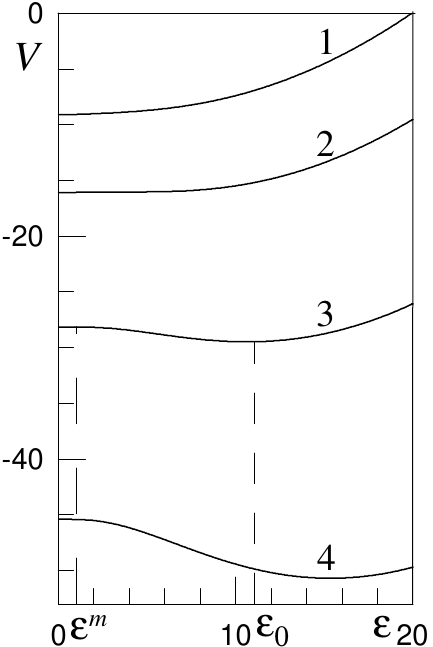}}
\caption{\label{fig1} Dependence of the synergetic potential on the strain at $g {=} 0.2,~\theta {=} \alpha {=} 0.25$ and various temperatures: (curve~1) $T_{\mathrm{e}} {<} T_{\mathrm{c}}^0$, (curve 2) $T_{\mathrm{e}}{=}T_{\mathrm{c}}^0$, (curve 3) $T_{\mathrm{c}}^0 {<} T_{\mathrm{e}} {<} T_{\mathrm{c}}$, and (curve 4) $T_{\mathrm{e}}{\geqslant } T_{\mathrm{c}0}$.}
\end{figure}
\begin{figure}[ht]
\centerline{\includegraphics[width=0.45\textwidth]{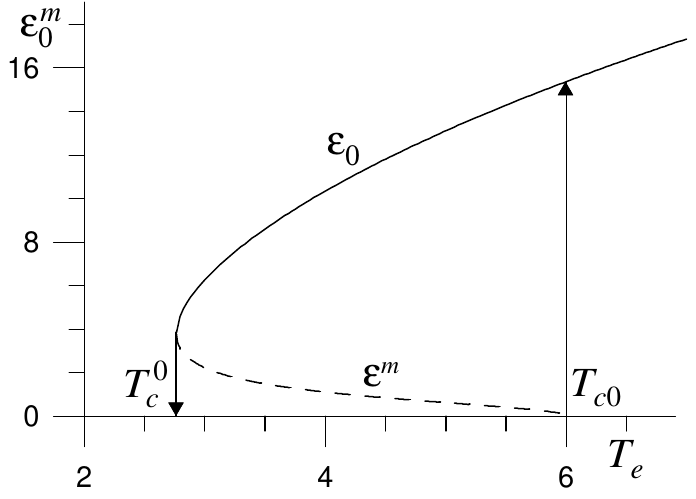}}
\caption{\label{fig2} Dependence of the steady-state values of the
strain on the temperature $T_{\mathrm{e}}$ at parameters of figure~\ref{fig1} (the solid curve corresponds to
the stable value $\varepsilon_0$, the dashed curve meets the unstable one,~$\varepsilon^{\mathrm{m}}$).}
\end{figure}
Within the adiabatic approximation (\ref{15}), the Lorenz equations (\ref{11})--(\ref{13}), where $G_\varepsilon$ is replaced by a dependence $G_\varepsilon (\varepsilon)$, is reduced to the Landau-Khalatnikov equation (\ref{18}). The synergetic potential has the form:
\begin{eqnarray}
V&=&\frac{1}{2}\varepsilon^{2} - \frac{g\alpha^2(T_{\mathrm{e}}-1)}{2} \left\{\frac{1}{g\alpha^2-\theta}
\ln\left| \frac{1+g\varepsilon^2}{1+\theta(\varepsilon/\alpha)^2} \right| \right. \nonumber \\
&&{}\left.+ \frac{1}{\theta g\alpha^2(\theta^{-1}-g^{-1}\alpha^{-2})} \left[ \theta^{-1} \ln\left| \theta^{-1}+(\varepsilon/\alpha)^2 \right|- g^{-1}\alpha^{-2}  \ln\left| g^{-1}\alpha^{-2} + (\varepsilon/\alpha)^2 \right| \right]
\right\}.
\label{51} \end{eqnarray}
Here, the constant $\alpha {\equiv} \varepsilon_{\mathrm{p}} / \varepsilon_s<1$ and the parameter $\theta {=} \Theta /G_\varepsilon<1$, describing the ratio of the tilts for the deformation curve on the plastic and the Hookean sections, are introduced. At a small value of temperature $T_{\mathrm{e}}$\,, the dependence~(\ref{51}) has a
monotonously increasing shape with its minimum at $\varepsilon = 0$ corresponding to the steady state of the absence of the adatom adsorption structure (curve~1 in figure~\ref{fig1}). As shown in figure~\ref{fig1}, at
\begin{equation} T_{\mathrm{c}}^0 {=} 1 {+} \theta g^{-1} {+} \alpha^{2} (1 {-} 2\theta) {+} 2\alpha \sqrt{ g^{-1}\theta (1{-}\theta)(1{-}g\alpha^{2})}, \label{52}
\end{equation}
a plateau appears (curve~2), which for $T_{\mathrm{e}} > T_{\mathrm{c}}^0$ is transformed
into a minimum, meeting the strain $\varepsilon_0\not= 0$,
and a maximum at $\varepsilon^{\mathrm{m}}$ that separates the minima corresponding to the
values $\varepsilon = 0$ and $\varepsilon = \varepsilon_0$ (curve~3) \cite{PRL_2006}. With
a further increase in the temperature $T_{\mathrm{e}}$, the ``ordered''
phase minimum, corresponding to the adatom adsorption configuration $\varepsilon = \varepsilon_0$, grows deeper, and the height of the interphase barrier
decreases vanishing at the critical value $T_{\mathrm{c}0}=1+g^{-1}$ (\ref{20}).
The steady-state values of the strain have the form (see figures~\ref{fig1} and \ref{fig2})
\begin{eqnarray}
\left( \varepsilon_0^{\mathrm{m}} \right)^2 = \left(2g\theta\right)^{-1} \left\{
g\left(T_{\mathrm{e}}-1-\alpha^{2} \right)-\theta {\mp} \sqrt{ \left[g\left(T_{\mathrm{e}}{-}1{-}\alpha^{2} \right){-}\theta \right]^2 {-} 4g\alpha^{2}\theta \left[1{-}g\left(T_{\mathrm{e}}{-}1 \right)\right] } \right\}, \label{55} \end{eqnarray}
where the lower sign meets the stable adatom structure and the upper sign corresponds to the unstable one. At $T_{\mathrm{e}}\geqslant  T_{\mathrm{c}0}$, the dependence $V (\varepsilon) $ is characteristic of the absence of the modulus defect (see curve~4 in figure~\ref{fig1}).

\begin{figure}[ht]
\centerline{\includegraphics[width=0.45\textwidth]{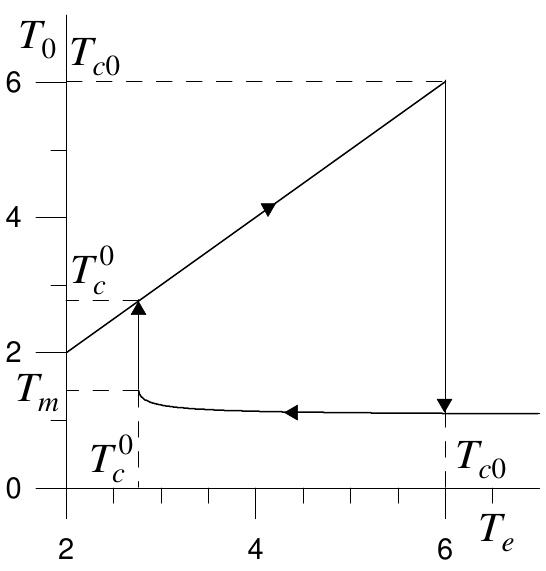}}
\caption{\label{fig3} Dependence of the steady-state value of the sample
temperature $T_0$ on the temperature $T_{\mathrm{e}}$ at parameters of figure~\ref{fig1}.}
\end{figure}

It is worth noting that the potential barrier inherent in the synergetic first-order transition manifests itself only due to the deformational defect of the modulus.
Since the latter is realized always \cite{PRL_2006}, it follows that the studied
adatom structure formation is a synergetic first-order transition. The considered situation differs from typical thermodynamic phase transitions. Really, in the latter case, the stationary value of the semiconductor temperature $T_0$ is equal to the thermostat value $T_{\mathrm{e}}$.
In this study, $T_0$ is reduced to the critical value $T_{\mathrm{c}0}$ for a
synergetic second-order transition (see section~\ref{sec:level3}). When the modulus defect is taken into account, the temperature
\begin {equation}
T_0=1+\frac{T_{\mathrm{e}} -1}{1+g\varepsilon_0^{2}}\,,
\label{56}
\end{equation}
whose value is defined by a minimum position of the
dependence (\ref{51}), is realized. In accordance with equations~(\ref{55}) and (\ref{56}), the
quantity $T_0$ monotonously decays from the value
\begin{eqnarray}
T_{\mathrm{m}}=1+\frac{T_{\mathrm{c}}^0 - 1}{1+g\left(\varepsilon_0^c\right)^{2}}\,, \qquad \varepsilon_0^c =  \left(2g\theta\right)^{-1} \left[
g\left(T_{\mathrm{c}}^0-1-\alpha^{2} \right)-\theta \right]
\label{58} \end{eqnarray}
at $T_{\mathrm{e}} = T_{\mathrm{c}}^0$, to $1$ at $T_{\mathrm{e}} \to \infty$.
As shown in figure~\ref{fig3}, the stationary temperature $T_0 = T_{\mathrm{e}}$ in the range from $0$ to $T_{\mathrm{c}0}$. The jump down occurs at $T_{\mathrm{e}}=T_{\mathrm{c}0}$, following which the value $T_{0}$ smoothly decreases. If the quantity $T_{\mathrm{e}}$ then decays, the steady-state temperature $T_0$ increases. At the point $T_{\mathrm{c}}^0$ [equation~(\ref{52})], the $T_0$ vs $T_{\mathrm{e}}$ dependence has a jump from $T_{\mathrm{m}}$ [equation~(\ref{58})] up to $T_{\mathrm{c}}^0$. For $T_{\mathrm{e}} < T_{\mathrm{c}}^0$, the steady-state temperature $T_0$ is also equal to $T_{\mathrm{e}}$.

Since $T_{\mathrm{c}}^0 > 1$, the maximal sample temperature (\ref{58}) is lower than the minimal temperature of the AFM tip (\ref{52}). As shown in figure~\ref{fig3}, at $T_{\mathrm{e}} > T_{\mathrm{c}}^0$, the stationary temperature $T_0$ of the sample is smaller than~$T_{\mathrm{e}}$.

\section{Summary}\label{sec:level5}

In accordance with the analysis presented above, the formation of an adatom adsorption structure
is caused by self-organization of shear components of the strain and by the stress fields,
on the one hand, and by the sample temperature, on the other hand. Here, the strain $\varepsilon$ acts as the order parameter, the conjugate field is reduced to the stress $\sigma$, and the temperature $T$ is the control parameter. The cause for self-organization is the positive feedback of $T$ and $\varepsilon$  on $\sigma$ [see equation~(\ref{12})]. According to equations~(\ref{6}) and
(\ref{8}), it is caused by the temperature dependence of the shear modulus. With an allowance for the effective shear modulus vs strain dependence, we obtain expressions for temperatures corresponding to the absolute instability of the adatom configuration $T_{\mathrm{c}}^0$ [equation~(\ref{52})] and its stability limit $T_{\mathrm{c}0}$ [equation~(\ref{20})]. A real thermodynamic transition temperature can be determined from the equality $V(0)=V(\varepsilon_0)$ of potentials in different phases and it is in the $(T_{\mathrm{c}}^0, T_{\mathrm{c}0})$ region. According to equation~(\ref{20}), systems predisposed to the formation of adatom structure have large typical $G_0$ and small effective $G_\varepsilon$ values of shear modulus.


The present study is principally different from \cite{KhYu}. In particular, the basic parameters and equations are different. For example, the order parameter is the strain, the shear modulus depends on temperature (\ref{8}), the derivation of equation for temperature (\ref{17gl}) differs and so on. Therefore, the resultant equations and figures are different from the ones in \cite{KhYu}. Thus, solid-liquid transition of an ultrathin lubricant film and self-organization of adatoms on the semiconductor surface in contact with the tip of the dynamic force microscope are described only by a similar approach but with many different aspects.

At a choice of real parameters, there is a difficulty connected primarily  with the following features which are common with the ultrathin lubricant film \cite{KhYu,TEPH-2010_1,jfd_2007,Khomenko2007165}. The properties of adatom layers due to their small thickness do not coincide with the properties of volume materials. They are characterized by various values of elastic constants, density, heat conductivity, etc. The adatom structure temperature is also an effective quantity, and it can essentially fluctuate, since the adatoms number is limited, and these fluctuations lead to transitions between states \cite{pla,TEPH_2005}. Therefore, we are restricted by a description of the qualitative system behavior and all parameters are transformed into dimensionless form.

\section*{Acknowledgements}

The basis of this study method was founded in the joint papers with my teacher Prof.~\mbox{A.I.~Olemskoi} cited here. I thank Dr. Bo N.J. Persson for the invitation, hospitality, helpful comments and suggestions during my stay in the Forschungszentrum J$\rm\ddot u$lich (Germany). I am grateful to him and to the organizers of the conference ``Joint ICTP-FANAS Conference on Trends in Nanotribology'' (12--16 September 2011, Miramare, Trieste, Italy) for the invitation and financial support for participation, during which this work was initiated. The work was supported by the grant of the Ministry of Education and Science of Ukraine ``Modelling of friction of metal nanoparticles and boundary liquid films which interact with atomically flat surfaces'' (No.~0112U001380) and by the grant for a research visit to the Forschungszentrum J$\rm\ddot u$lich (Germany). I am thankful to Dr.~Boris Lorenz for an attentive reading and correction of this article.

%
%
\newpage
\ukrainianpart

\title{Самоорганізація структури адсорбованих адатомів при взаємодії із зондом динамічного силового мікроскопа}

\author{О.В. Хоменко\refaddr{label1,label2}}
\addresses{
\addr{label1} Кафедра моделювання складних систем, Сумський державний університет, вул. Римського-Корсакова, 2, 40007 Суми, Україна
\addr{label2} Інститут Петера Грюнберга-1, Дослідницький центр Юліха, 52425 Юліх, Німеччина
}

\makeukrtitle

\vspace{-0.25cm}
\begin{abstract}
\tolerance=3000%
Формування структури адсорбованих адатомів при дослідженні в режимі динамічної силової мікроскопії представлено як результат спонтанної появи зсувної деформації в результаті зовнішнього над\-кри\-тич\-но\-го нагрівання. Цей перехід описується рівнянням Кельвіна-Фойгта для в'язкопружного середовища, релаксаційним рівнянням Ландау-Халатнікова для зсувних напружень та релаксаційним рівнянням для температури. Показано, що ці рівняння формально збігаються із синергетичною системою Лоренца, де зсувна деформація відіграє роль параметра порядку, спряжене поле зводиться до напружень, та температура є керувальним параметром. В рамках адіабатичного наближення знайдені стаціонарні значення цих величин.
Враховуючи залежність модуля зсуву зразка від деформації, формування конфігурації адсорбованих адатомів описано як перехід першого роду. Критична температура зонда лінійно зростає з ростом ефективного значення модуля зсуву зразка і зменшується при зростанні його характерного значення.
\keywords фазовий перехід, реологія, пластичність, деформація, напруження, атомно-силова мікроскопія
\end{abstract}

\end{document}